\documentclass[a4paper,conference]{IEEEtran}
\usepackage[dvipdfm]{graphicx}
\usepackage[dvipdfm]{color}
\usepackage{bm} 
\usepackage[colorlinks=true,dvipdfm]{hyperref}

\begin{document}

\title{Roles of hubs in Boolean networks}

\author{\authorblockN{Chikoo Oosawa}
\authorblockA{
Department of Bioscience and Bioinformatics, Kyushu Institute of Technology\\
Email: chikoo@bio.kyutech.ac.jp
}
}

\maketitle


\begin{abstract}
We examined the effects of inhomogeneity on the dynamics and structural properties using Boolean networks. Two different power-law rank outdegree distributions were embedded to determine the role of hubs. The degree of randomness and coherence of the binary sequence in the networks were measured by entropy and mutual information, depending on the number of outdegrees and types of Boolean functions for the hub. With a large number of outdegrees, the path length from the hub reduces as well as the effects of Boolean function on the hub are more prominent. These results indicate that the hubs play important roles in networks' dynamics and structural properties. By comparing the effect of the skewness of the two different power-law rank distributions, we found that networks with more uniform distribution exhibit shorter average path length and higher event probability of coherence but lower degree of coherence. Networks with more skewed rank distribution have complementary properties. These results indicate that highly connected hubs provide an effective route for propagating their signals to the entire network.
\end{abstract}

\IEEEpeerreviewmaketitle

\section{Introduction}
Recent studies of complex networks \cite{AB02} have shown inhomogeneous connectivities having a small number of highly connected nodes, frequently called as hubs, along with many poorly connected nodes. The inhomogeneity in connectivity has large effects on the property and/or function of the hubs; for example robustness in metabolic, genetic regulatory and neural networks \cite{Book2006}; performance of artificial neural networks \cite{HF03}; and dynamics of coupled oscillators \cite{NISHI03,ZARA06}.

The Boolean network \cite{ORIGIN} is one of the discrete dynamical models for the transcriptional regulatory network and exhibits binary sequences of the state variables that represent expression pattern of the network \cite{ORIGIN,COMAS}. Since the state variables in the network are sensitive to inputs from other nodes via directed edges, and affect other nodes, the quality of communication is characterized as the size of mutual information \cite{SHA63}. This mutual information indicates the degree of coherence, synchronization, amount of information content in the state variables, or potential for computational capability of the network \cite{LANGTON}.

In this study, we show the role of hubs for the emergence of coherence in a Boolean network. Since we embedded power-law rank outdegree distributions in the Boolean networks with an input connectivity of $K_{in}$= 2, the model networks have some hubs that integrate many outdegree connections. Because the hubs synchronously transmit their state to the downstream nodes, these nodes are simultaneously affected by single or multiple hubs . The structural condition seems to automatically provide global coherence in the state variables; however, the structural aspects give only the possible effects of the hubs. In fact, we need to consider a type of Boolean function at the hubs and path length from the hubs. We show both the effects of Boolean functions and path length on both the event probability and size of entropy and mutual information.

\section{Model}
Dynamics of the Boolean networks \cite{ORIGIN,COMAS} are determined by 
\begin{equation}
X_i(t+1)=B_i\left[\bm{X}(t)\right]\quad(i=1,2,...,N),
\label{eq:bn}
\end{equation}
where $X_{i}(t)$ is the binary state, 0 or 1, of node {\it i} at time $t$; $B_i(\cdot)$ is the Boolean function [see Tables \ref{tab:boolA}--\ref{tab:boolC}] used to simultaneously update the state of node {\it i}; and $\bm{X}(t)$ is a binary vector that gives the states of the $N$ nodes in the network [See Fig. \ref{fig:3nodes}(a)]. After assigning initial states $\bm{X}(0)$ to the nodes, the successive states of the nodes are updated by input states and their Boolean function. The dynamical behavior of these networks is represented by the time series of the binary states. The time course follows a transient phase from an initial state until a periodic pattern, known as an attractor, is established.

\begin{table}[h]
\caption{AND-OR type Boolean functions with indegree $K_{in}$= 2. Since output probability of one or zero equals $\frac{1}{4}$ or $\frac{3}{4}$  with random binary inputs the functions have the same output entropy, $H\left(\frac{1}{4}\right)=H\left(\frac{3}{4}\right)=0.811$.}
\begin{center}
\begin{tabular}{|cc|*{8}{r}|}\hline
\multicolumn{2}{|c|}{Input}&\multicolumn{8}{c|}{Output}\\ \hline
0&0&0&0&0&1&0&1&1&1\\
0&1&0&0&1&0&1&0&1&1\\
1&0&0&1&0&0&1&1&0&1\\
1&1&1&0&0&0&1&1&1&0\\
\hline
\end{tabular}\label{tab:boolA}
\end{center}
\end{table}

\begin{table}[h]
\caption{XOR type Boolean functions with indegree $K_{in}$= 2. Since output probability of one or zero equals $\frac{1}{2}$ with random binary inputs the functions have the same output entropy, $H\left(\frac{1}{2}\right)=1$.}
\begin{center}
\begin{tabular}{|cc|*{6}{r}|}\hline
\multicolumn{2}{|c|}{Input}&\multicolumn{6}{c|}{Output}\\ \hline
0&0&0&0&0&1&1&1\\
0&1&0&1&1&0&0&1\\
1&0&1&0&1&0&1&0\\
1&1&1&1&0&1&0&0\\
\hline
\end{tabular}\label{tab:boolX}
\end{center}
\end{table}

\begin{table}[h]
\caption{CONSTANT type Boolean functions with indegree $K_{in}$= 2. Since output probability of one or zero equals 1 or 0 with random binary inputs the functions have the same output entropy, $H(0)=H(1)=0$.}
\begin{center}
\begin{tabular}{|cc|*{2}{r}|}\hline
\multicolumn{2}{|c|}{Input}&\multicolumn{2}{c|}{Output}\\ \hline
0&0&0&1\\
0&1&0&1\\
1&0&0&1\\
1&1&0&1\\
\hline
\end{tabular}\label{tab:boolC}
\end{center}
\end{table}

\section{Numerical Condition}
\begin{figure}[h]
\begin{center}
\includegraphics[scale=0.35]{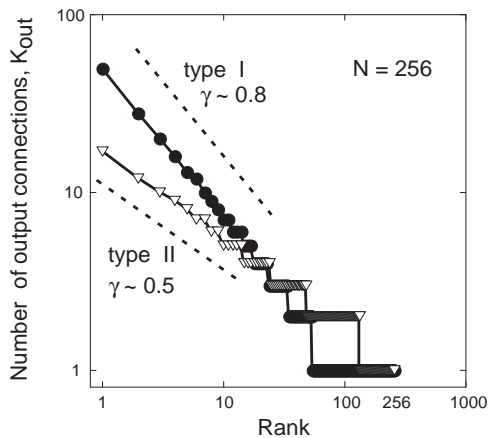}
\caption{Power-law rank connectivity distributions in the model. Power exponent, $\gamma$ for type I and type II is about 0.8 and 0.5, respectively, where $K_{out}(rank) \sim rank^{-\gamma}$. We performed only a single network size of 256 ($=N$) in this paper.}
\label{fig:topo}
\end{center}
\end{figure}

\begin{figure}[h]
\begin{center}
\includegraphics[scale=0.45]{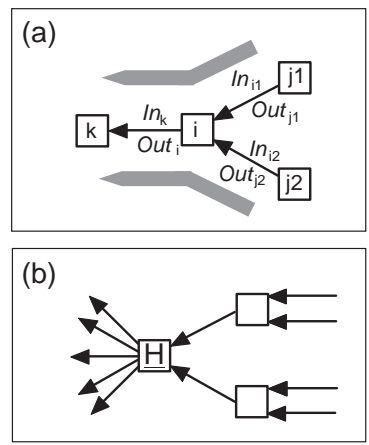}
\caption{(a) Flow of state variables from upstream to downstream. Input sequence $In_{i1}$ for node $i$ is the same as the output sequence of an upstream node $Out_{j1}$, and the output sequence $Out_i$ for node $i$ is the same as the input sequence of a downstream node $In_k$. When node $i$ has multiple output connections, it has the same binary sequence because state variables in networks are subject to Eq. (\ref{eq:bn}). (b) Typical local structure around a hub. Squares and directed edges (arrows) correspond to the nodes and connections of binary sequence pathways, respectively. \underline{H} in the squares corresponds to the Boolean function. Each node has one of the Boolean functions as shown in Tables \ref{tab:boolA}--\ref{tab:boolC} [see also Section \ref{sec:appen}].}
\label{fig:3nodes}
\end{center}
\end{figure}

We randomly constructed $10^4$ Boolean networks in each power-law rank distribution [see Fig. \ref{fig:topo}] with a fixed network size. 2 $\times 10^3$ initial states were applied to each network. Sixteen different Boolean functions [see Tables \ref{tab:boolA}--\ref{tab:boolC}] were used with equal probabilities. Note that all the generated networks use the same amount of resources since the size of the network is fixed--256 nodes, 512 directed edges, and 16 sets (16$^2$ = 256) for all the Boolean functions [see Table \ref{tab:frqBF} in Section \ref{sec:appen}].
We measured entropy (randomness) and mutual information (coherence) of the state variables to characterize the dynamics of the Boolean networks \cite{COMAS,SHA63,LANGTON} [see Fig. \ref{fig:3nodes}].
\section{Results}
\subsection{Dynamics}
In total, we obtained 137254 and 459240 attractors from type I and type II distributions, respectively. The size of entropy and mutual information were measured from the attractors. Major statistics of dynamical properties are shown in Table \ref{tab:dy}. By comparing the effect of the skewness of the two different power-law rank distributions, we found that on average networks with more uniform distribution (type II) exhibit higher event probability of coherence, but lower degree of coherence.
\begin{table}[h]
\caption{Dependence of dynamical properties on outdegree distributions: Magnitude relations are indicated.}
\begin{center}
\begin{tabular}{|cc|ccc|}
\hline
Properties&&Type I&&Type II\\
\hline
No. of attractors&from $10^{4}$ nets&137254&$<$&459240\\
\hline
Proportion of&positive entropy&0.788&$<$&0.848\\
Proportion of&positive mutual information&0.661&$<$&0.698\\
\hline
&Median&34.8&$>$&31.6\\
Entropy / bit&Average&46.1&$>$&38.4\\
&3rd quartile&83.3&$>$&63.2\\
\hline
Mutual&Median&0.827&$>$&0.387\\
information / bit&Average&2.48&$>$&0.996\\
&3rd quartile&3.73&$>$&1.36\\
\hline
\end{tabular}\label{tab:dy}
\end{center}
\end{table}
\subsection{Path length}
\begin{figure}[htb]
\begin{center}
\includegraphics[scale=0.3]{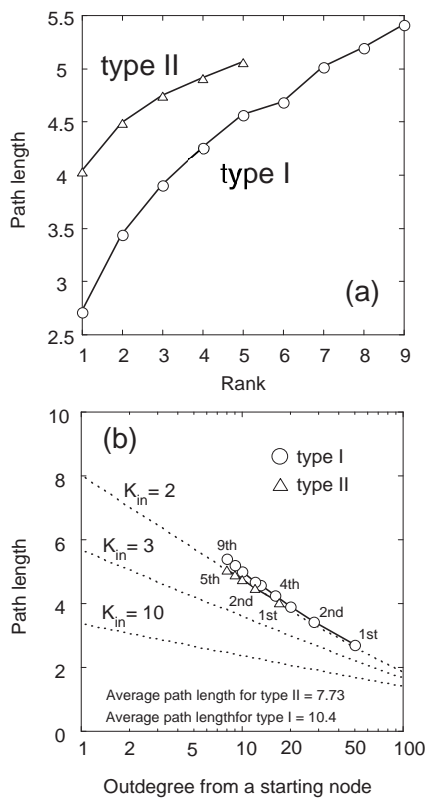}
\caption{(a) Relationships between the rank of the hubs [see Fig. \ref{fig:topo}] and the path length. (b) Dependence of path length on the number of outdegree from a starting node. The dotted lines show the relationship of Eq. (\ref{eq:steps}) [see Section \ref{sec:appen}]. Both path lengths are obtained from $10^{4}$ generated networks with $N=$ 256.}
\label{fig:pathlen}
\end{center}
\end{figure}

To get stractual property of propagating route of state variables, we measured two properties: 
\begin{enumerate}
\item Path length which the average number of the directed edges in the shortest path from a node to all reachable nodes.
\item Average path length which the average number of the path lengths for all the nodes.
\end{enumerate}
Figure \ref{fig:pathlen} shows the differences in the both path lengths. Nodes with higher outdegrees have shorter path lengths.

Theoretical relationship between the number of outdegrees, $X$ of a starting node and path length, $i$ can be written as [For details, see Section \ref{sec:appen}.]
\begin{equation}
i=Log_{K_{in}}\left(\frac{255}{X}+1\right).
\label{eq:steps}
\end{equation}
We obtained good relationship between numeical results and Eq. (\ref{eq:steps}).
\subsection{Rank Dependent Dynamics}
\begin{figure}[hbt]
\begin{center}
\includegraphics[scale=0.30]{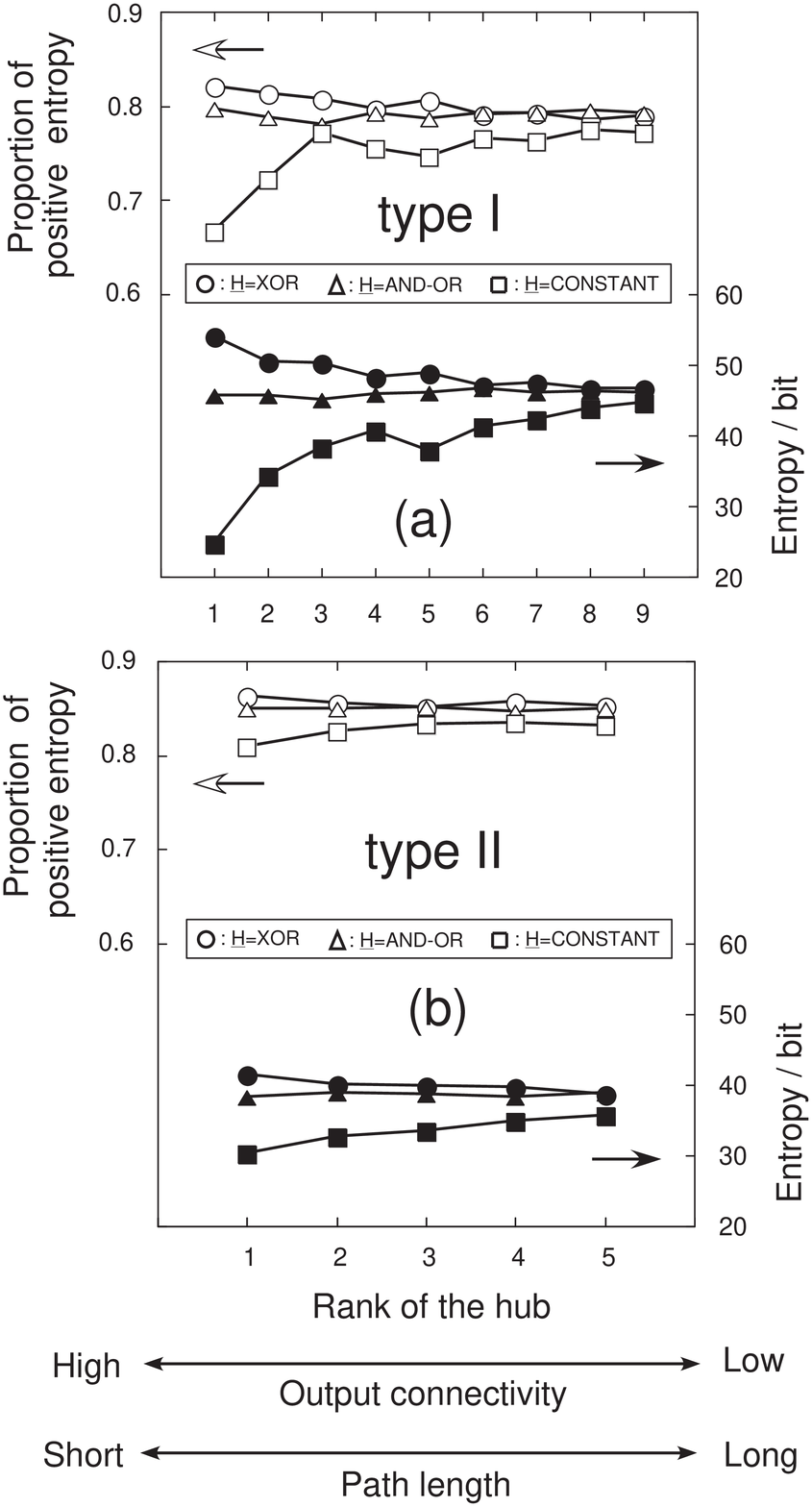}
\caption{Dependence of proportion of positive entropy and size of entropy (a) and (b) on type I and type II distribution in Fig. \ref{fig:topo}. Three different symbols denote different types of Boolean functions of the hubs: square; CONSTANT type; triangle; AND-OR type; and circle; XOR type [see Tables \ref{tab:boolA}--\ref{tab:boolC}]. \underline{H} denotes the Boolean functions of the hubs [see Fig. \ref{fig:3nodes} and Table \ref{tab:frqBF} in Section \ref{sec:appen}].}
\label{fig:rankEn}
\end{center}
\end{figure}

\begin{figure}[hbt]
\begin{center}
\includegraphics[scale=0.30]{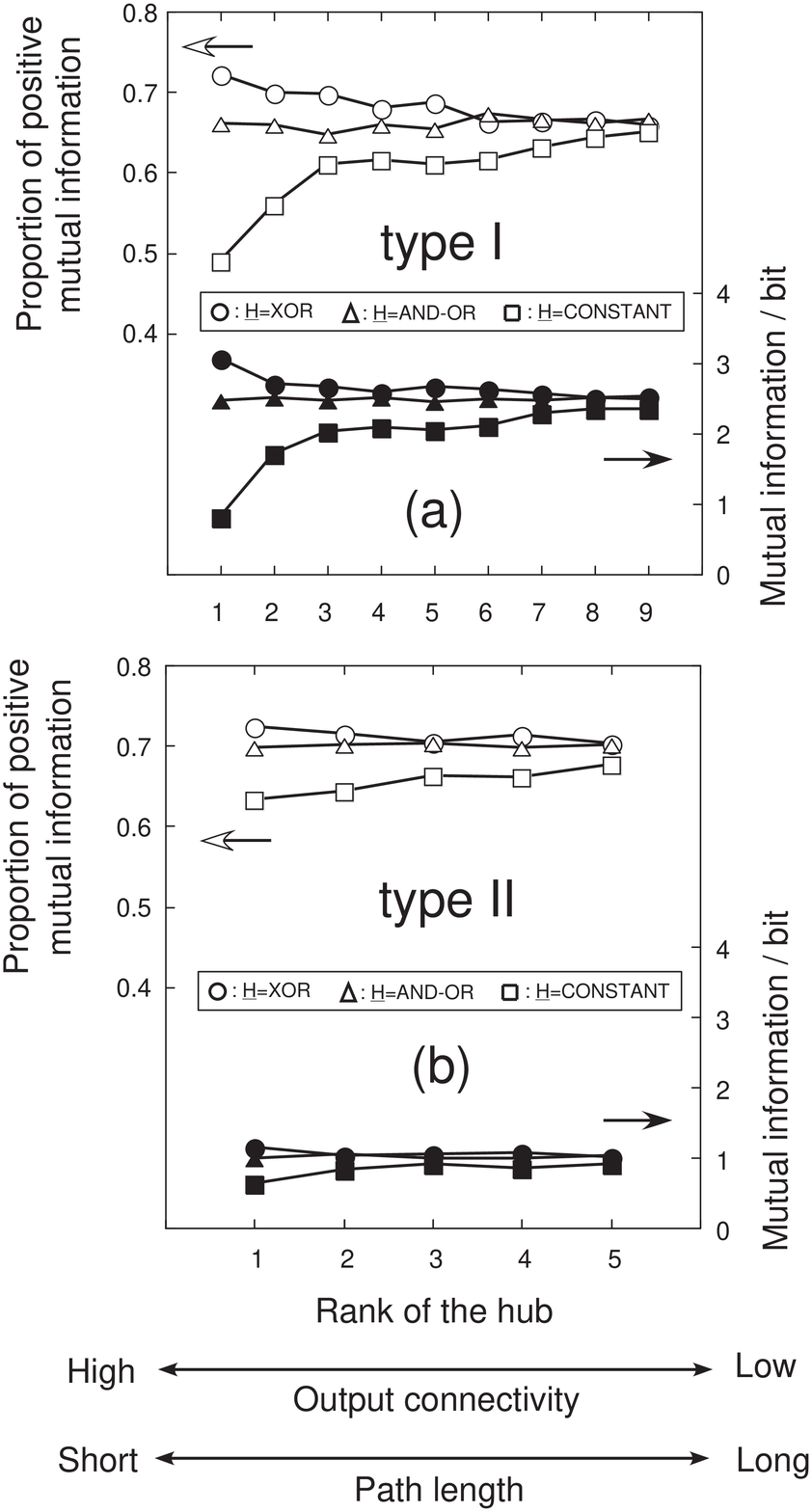}
\caption{Dependence of proportion of positive mutual information and size of mutual information (a) and (b) on type I and type II distribution in Fig. \ref{fig:topo}. Three different symbols denote different types of Boolean functions on the hubs: square; CONSTANT type; triangle; AND-OR type; and circle; XOR type [see Tables \ref{tab:boolA}--\ref{tab:boolC}]. \underline{H} denotes the Boolean functions of the hubs [see Fig. \ref{fig:3nodes} and Table \ref{tab:frqBF} in Section \ref{sec:appen}].}
\label{fig:rankMi}
\end{center}
\end{figure}

Together with the network structural condition, we show the dependence of rank distributions on entropy and mutual information in Figs. \ref{fig:rankEn} and \ref{fig:rankMi}. We classified different 16 Boolean functions into three types (AND-OR, XOR, and CONSTANT types) based on the input--output relationships \cite{EIC05a,AROB07}.
The dependence of the Boolean function on entropy and mutual information is prominent in higher ranked hubs on both the outdegree distribution styles. The Boolean functions with larger entropy tend to have larger mutual information. These results have weak dependence but clear tendencies, and suggest that the collective (global) coherence in the state variables of the networks is subject to the style of upstream (local) conditions, including the number of outdegrees from a hub and assignment of the Boolean functions.
\section{Conclusions}
We have summarized the results in Table \ref{tab:summary}. A critical control parameter for synchronizability of coupled oscillators in complex networks is coupling strength or path length among the oscillators \cite{NISHI03,ZARA06}. Since our Boolean networks consist of the same amount of resources, some nodes integrate many edges, while others have less. As shown in Fig. \ref{fig:pathlen}, with a larger number of outdegree from hubs, the path length from the hubs reduces. The shorter path length can result in a stronger coupling strength between the hubs and other nodes. Because of the differences between type I and type II distributions, the higher ranked hub in type I distribution have an advantage to transmit their signals [see Figs. \ref{fig:rankEn}(a) and \ref{fig:rankMi}(a)], and the hubs in type II distribution have longer path lengths and show smaller dependences [see Figs. \ref{fig:rankEn}(b) and \ref{fig:rankMi}(b)]. On the other hand, type II distributions exhibit higher event probability of entropy and mutual information and a large number of attractors [see Tables \ref{tab:dy} and \ref{tab:summary}]. The resultant dynamical properties are due to the more decentralized topology that can lead to the presence of some mid-rage hubs. The hubs permit local and/or weak coherences to emerge in the networks.

The higher indegree $K_{in}>2$ provides higher density of the edges as well as a narrower range of the path length [see Fig. \ref{fig:pathlen}], and therefore the dependence of the outdegree from a hub will become less prominent.
\begin{table}[hbt]
\caption{Summary of results: Magnitude relations are indicated.}
\begin{center}
\begin{tabular}{|c|c|}
\hline
Dynamical and structural properties&Types\\
\hline
\hline
Boolean function dependence&XOR$>$AND-OR$>$CONSTANT\\
\hline
Skewness of outdegree distribution&I\quad $>$\quad II\\
\hline
Average path length&I\quad $>$\quad II\\
\hline
Size of entropy and mutual information&I\quad $>$\quad II\\
\hline
Rank dependence&I\quad $>$\quad II\\
\hline
Event probability of positive&\\
entropy and mutual information&I\quad $<$\quad II\\
\hline
\end{tabular}\label{tab:summary}
\end{center}
\end{table}

\subsection*{Acknowledgments}
This work was supported by a Grant-in-Aid No. 18740237 from MEXT (JAPAN).
\section{Appendix}
\label{sec:appen}
Since the networks are randomly constructed with $K_{in}=2$, all the nodes have two downstream nodes, on average, regardless of outdegree distribution styles. The total number of downstream nodes from a starting node $T(i)$ can be written as
\begin{equation}
T(i)=\sum_{n=0}^{i}2^{n}=2^{i}-1,
\end{equation}
where $i$ is the number of steps (path length) from the starting node. With increasing $i$, the total number of downsteam nodes increases. When the starting node has $X$ outdegree,
\begin{equation}
XT(i)=X(2^{i}-1).
\end{equation}
Since the size of network is limited to 256 in the paper, the total number will reach to the 255.
\begin{equation}
X(2^{i}-1)=255,
\end{equation}
where 255 indicates the exclusion of a starting node from the network.
Finally, we can obtain the number of steps (path length),
\begin{equation}
i=Log_{2}\left(\frac{255}{X}+1\right).
\label{eq:last}
\end{equation}
When $X = 1$, the $i$ takes 8, meaning that 8 steps are required to reach other nodes in the same network on average [see Fig. \ref{fig:pathlen}(b)].
For general form of Eq.(\ref{eq:last}),
\begin{equation}
i=Log_{K_{in}}\left(\frac{(N-1)(K_{in}-1)}{X}+1\right),
\end{equation}
where $K_{in}$ is indegree of the Boolean network and N is the number of nodes in the network.
\\
\begin{table}[htb]
\caption{Number of networks out of $10^{4}$ networks (type I) are indicated. Three different types of Boolean functions [see Tables \ref{tab:boolA}--\ref{tab:boolC}] at the hubs are used with equal probability. "\underline{H}" in the table denotes the hubs Boolean function [see Fig.\ref{fig:3nodes}(b)].}
\begin{center}
\begin{tabular}{|c||c|c|c|}
\hline
Rank&\underline{H}=XOR&\underline{H} = AND-OR&\underline{H}=CONSTANT\\
\hline
1&3725&4995&1280\\
2&3736&5042&1222\\
3&3696&5029&1275\\
4&3710&5020&1270\\
5&3705&5095&1200\\
6&3725&4996&1279\\
7&3756&5006&1238\\
8&3768&5012&1220\\
9&3815&4951&1234\\
\hline
\end{tabular}\label{tab:frqBF}
\end{center}
\end{table}

\end{document}